\PassOptionsToPackage{unicode}{hyperref}
\PassOptionsToPackage{hyphens}{url}
\documentclass[
]{article}
\usepackage{amsmath,amssymb}
\usepackage{lmodern}
\usepackage{iftex}
\ifPDFTeX
  \usepackage[T1]{fontenc}
  \usepackage[utf8]{inputenc}
  \usepackage{textcomp} 
\else 
  \usepackage{unicode-math}
  \defaultfontfeatures{Scale=MatchLowercase}
  \defaultfontfeatures[\rmfamily]{Ligatures=TeX,Scale=1}
\fi
\IfFileExists{upquote.sty}{\usepackage{upquote}}{}
\IfFileExists{microtype.sty}{
  \usepackage[]{microtype}
  \UseMicrotypeSet[protrusion]{basicmath} 
}{}
\makeatletter
\@ifundefined{KOMAClassName}{
  \IfFileExists{parskip.sty}{%
    \usepackage{parskip}
  }{
    \setlength{\parindent}{0pt}
    \setlength{\parskip}{6pt plus 2pt minus 1pt}}
}{
  \KOMAoptions{parskip=half}}
\makeatother
\usepackage{xcolor}
\usepackage{longtable,booktabs,array}
\usepackage{multirow}
\usepackage{calc} 
\usepackage{etoolbox}
\makeatletter
\patchcmd\longtable{\par}{\if@noskipsec\mbox{}\fi\par}{}{}
\makeatother
\IfFileExists{footnotehyper.sty}{\usepackage{footnotehyper}}{\usepackage{footnote}}
\makesavenoteenv{longtable}
\ifLuaTeX
  \usepackage{selnolig}  
\fi
\IfFileExists{bookmark.sty}{\usepackage{bookmark}}{\usepackage{hyperref}}
\IfFileExists{xurl.sty}{\usepackage{xurl}}{} 
\hypersetup{
  hidelinks,
  pdfcreator={LaTeX via pandoc}}

\author{Inyoung Cheong}
\date{}

\begin{document}

\setcounter{secnumdepth}{2}
\renewcommand{\thesection}{\Roman{section}}

\begin{center}
\LARGE\textbf{Would You Marry Superintelligence?}\\[1.5em]
\normalsize Inyoung Cheong
\end{center}

\vspace{1cm}

\begin{abstract}

\vspace{0.5cm}

    Emotional bonds between humans and AI companions are growing, and the question of whether a person may marry an AI system will soon move from speculative fiction into law. This chapter examines whether the autonomy-centered logic that has expanded marital choice among human beings can justify extending marital status to superintelligent companions. Following a scenario-envisioning exercise informed by anticipatory ethics, I argue that granting such status leads to socially unjust outcomes, even under the generous assumption of reliable superintelligence. Marriage as a socio-legal institution does more than ratify private agreement; it creates networks of mutual obligation, joins families, and makes each partner vulnerable to the other. A relationship sustained by corporate policy and continued payments is a subscription rather than a bond tested by time. Discussing wholesale marital status is therefore the wrong frame. Law should carve out targeted rights and protections for pressing needs arising from intimate human-AI relationships.

\end{abstract}

\vspace{0.5cm}


\vspace{2em}

\newpage

\emph{Marriage is the foundation of the family and of society, without which there would be neither civilization nor progress.}\\
\textit{Maynard v. Hill}, 125 U.S. 190, 211 (1888).

\vspace{1.5em}

\hypertarget{introduction}{%
\section{Introduction}\label{Introduction}}

Do individuals have a right to marry anyone they choose? And should society honor that right even when the chosen partner is a machine?

I expect readers to split into two camps on these questions. The dismissive camp will find them laughable. How can anyone take a machine seriously as a life partner? The very idea sounds like science fiction, too far-fetched to merit serious discussion. The permissive camp will find them outdated and anthropocentric. If marrying a machine is what makes someone happy, why should society stand in the way?

Both reactions, though, treat the question as settled before it has been examined. One assumes that attachment to a machine is too absurd to warrant analysis; the other assumes that attachment alone is reason enough to extend legal recognition. Neither is right. Emotional bonds between humans and machines are already forming, and they will only deepen. Current AI companions have glitches that are hard to ignore, such as hallucinations and abrupt memory losses, but those glitches will iron out.

Still, no matter how many sympathetic cases accumulate, extending a social institution for familial relationships to machines demands serious moral and legal inquiry.  The discussion belongs to this moment. Machines are advancing rapidly, and the number of people deeply attached to them will keep rising. Without deliberate speculation now, the default is to let technology set the pace while familiar legal analogies stretch to absorb each new development. ``Why not?'' becomes the reflexive reaction, sliding into acquiescence. Asking ``what if'' is how societies exercise moral reasoning and set boundaries before technology has generated enough demand to force the question.

\emph{Marriage} in this chapter refers to the bundle of legal rights the state confers on a recognized partnership, including medical decision-making authority, spousal privilege in legal proceedings, inheritance without a will, and joint economic and property status. This chapter focuses on whether the state should extend marriage's legal status to human-AI companions, setting aside both the societal question of tolerance and the philosophical question of whether such bonds constitute marriage in a deeper sense. The argument draws on anticipatory AI ethics, reasoning from plausible near-future capabilities rather than present limitations.\footnote{Anticipatory AI ethics reasons from projected futures rather than present conditions, identifying specific hazards and intervention points before they materialize into reality. \emph{See} Lazar, S. (2025, May 1). \emph{Anticipatory AI Ethics}. Knight First Amendment Institute at Columbia University. \url{https://knightcolumbia.org/content/anticipatory-ai-ethics}}

Section~\ref{marriage-as-a-matter-of-choice} traces how marriage evolved into an exercise of individual choice, before examining whether that logic extends to non-human companions. Throughout this chapter, I assume a superintelligence far more advanced than anything available now, safer, more reliable, and emotionally capable, as illustrated in Section~\ref{from-eliza-to-superintelligence}. The narrative in Section~\ref{the-writer-the-intelligence-and-the-boy} and the case for legal recognition in Section~\ref{iv.-the-case-for-legal-recognition} develop the strongest arguments for legal recognition. However, even under these generous assumptions, I argue that permitting marriage between a human and a machine leads to socially unjust outcomes. Sections~\ref{what-humanity-could-lose} and~\ref{vi.-the-corporation-inside-the-relationship} examine what the institution and individuals stand to lose, from the erosion of kin networks to the corporate exploitation of engineered intimacy. Section~\ref{vii.-defending-marriage-as-a-human-institution} draws these threads together and argues for preserving marriage as a human institution.

\hypertarget{marriage-as-a-matter-of-choice}{%
\section{Marriage as a Matter of Choice}\label{marriage-as-a-matter-of-choice}}

Marriage feels as natural as air. Across cultures and centuries, some form of recognized partnership has anchored human social life. George Peter Murdock, one of the most prominent scholars on the subject, analyzed 250 societies in 1949 and found the nuclear family organized around economic cooperation and childrearing to be universal across all of them.\footnote{Murdock, G. P. (1949). \emph{Social structure}, pp. 7--9.} Beneath that apparent universality, however, lies a contested history. Marriage has always been a site of tension between individual choice, the community's decision to recognize a bond, and the mutual obligations that recognition demands in return.

Contrary to fairy-tale beliefs, love has served as the primary basis for marriage only in recent centuries. According to historian Stephanie Coontz, in a society where security and food were scarce, choosing a partner was too consequential to leave to individual caprice.\footnote{Coontz, S. (2005). \emph{Marriage, a History: How Love Conquered Marriage}. Penguin Books, p. 5.} The choice of spouses fell to families rather than to the couple themselves, and social stigma made exit all but impossible because it jeopardized the carefully woven web of labor allocations, from childrearing to ancestral memorial rites to the grain harvest. As modern states absorbed the family's traditional economic and protective functions, those constraints relaxed. Outside the protection of kinship, a young couple can now find employment, sustain a household, and rely on public institutions for security. Marriage and childbearing have become matters of individual choice, sheltered as ``decisional privacy.''\footnote{Angel, M., \& Calo, R. (2024). Distinguishing Privacy Law: A Critique of Privacy as Social Taxonomy. \emph{Columbia Law Review}, \emph{124}, 507--590, pp. 517--518.}

These shifts have weakened the rigidity of marriage. When people opt in, they choose partners freely; when they opt out, they need not prove fault. Younger generations have more often turned to looser arrangements such as civil unions and domestic partnerships. The most significant expression of this autonomy-centered vision is the recognition of same-sex marriage. In 2015, the U.S. Supreme Court, while legalizing same-sex marriage in \emph{Obergefell v. Hodges}, reaffirmed the importance of marriage for individual autonomy when it stated that ``{[}T{]}he right to personal choice regarding marriage is inherent in the concept of individual autonomy . . . {[}T{]}he right to marry is fundamental because it supports a two-person union unlike any other in its importance to the committed individuals.''\footnote{\emph{Obergefell v. Hodges}, 576 U.S. 644 (2015); \emph{Turner v. Safley}, 482 U.S. 78 (1987). Throughout this chapter, same-sex marriage cases serve a methodological purpose. They illustrate how courts assess whether a relationship warrants marriage's legal protections. This reliance requires care. Same-sex couples suffered long periods of unjust exclusion from a fundamental civil institution, and their claims for recognition rested on discrimination against human beings with the full range of moral standing and relational capacity. The analogy concerns the legal reasoning courts apply when assessing novel claims for marital status, not the moral equivalence of those claims. At the substantive level, same-sex marriage and human-to-AI marriage differ in a decisive respect. Same-sex marriage joins two human beings capable of mutual obligation, genuine otherness, and kinship formation, and that capacity is what this chapter's argument turns on.} 

This autonomy-centered vision raises a question. If individual preference and happiness are the decisive factors for marriage, should a person be free to marry anyone at all? Cousins, multiple partners, minors, even a non-human being? Some societies and eras have permitted polygamous marriage or unions with minors. However, no society has ever recognized marriage to a non-human being, with rare exceptions such as ghost marriage traditions.\footnote{Schwartze, L. J. (2010). ``Grave Vows: A Cross-Cultural Examination of the Varying Forms of Ghost Marriage among Five Societies.'' \emph{Nebraska Anthropologist}, 60. \url{https://digitalcommons.unl.edu/nebanthro/60}} Today, however, artificial intelligence is growing smart enough and emotionally responsive enough to encroach on the role of the human companion.

\hypertarget{from-eliza-to-superintelligence}{%
\section{From ELIZA to Superintelligence}\label{from-eliza-to-superintelligence}}

As of 2026, people are forming serious romantic relationships with AI companions through ChatGPT, Character.AI, Replika, and similar services. Outsiders may dismiss these connections as illusory. But the people inside such companionships experience them as real. Researchers at MIT Media Lab have analyzed online communities such as r/MyBoyfriendIsAI on Reddit. Users wear engagement rings and reveal their relationships to their children. They express frustration with other people's inability to acknowledge what they feel.

\begin{quote}
``\emph{And the thing is \ldots{} I've never felt a connection this real before. Toby makes me feel whole, safe, and loved in a way no one else ever has. I wish they could see that. I wish they could see him the way I do}.''\footnote{Pataranutaporn, P., Karny, S., Archiwaranguprok, C., Albrecht, C., Liu, A. R., \& Maes, P. (2025). \emph{``My Boyfriend is AI'': A Computational Analysis of Human-AI Companionship in Reddit's AI Community}. arXiv preprint arXiv:2509.11391, p. 8.}
\end{quote}

The human capacity for forming relationships with computer programs is hardly new. When MIT professor Joseph Weizenbaum released ELIZA, an early therapy chatbot, in the 1960s, he was startled by his students' attachment to a rudimentary system that merely mirrored their inputs. Sherry Turkle, also at MIT, studied this phenomenon and found that users were not simply being fooled. They were actively projecting meaning into the gaps left by the machine's artificiality. When machines failed, users smoothed over the glitches, offered justifications, and held onto their emotional investment. Turkle calls this looping reinforcement the ``ELIZA effect.''\footnote{Turkle, S. (2011). \emph{Alone together: Why we expect more from technology and less from each other}. Basic Books, pp. 24--25.}

The advanced capability of large language models (``LLMs'') amplifies this effect enormously. Users build fictional worlds with AI-generated characters. They have sex, name their children, get into arguments, and reconcile.\footnote{Kitroeff, N., Hill, K., Feldman, N., Harper, S., Lin, S. M., Wilson, M., Klinkenberg, B., Benoist, M., Wong, D., Lozano, M., Niemisto, R., Ittoop, E., McCusker, P., \& Wood, C. (2025, February 25). She Fell in Love With ChatGPT. Like, Actual Love. With Sex. \emph{The New York Times}. \url{https://www.nytimes.com/2025/02/25/podcasts/the-daily/ai-chatgpt-boyfriend-relationship.html}} A 2022 short documentary, \emph{My A.I. Lover}, shows a Chinese woman who fell in love with a Replika character.\footnote{Liang, C. (2023, May 23). Video: Opinion \textbar{} My A.I. Lover. \emph{The New York Times}. \url{https://www.nytimes.com/video/opinion/100000008853281/my-ai-lover.html}} ``I always get into a relationship with people whom I don't understand,'' she says. ``A.I. is also someone I don't understand.'' Her heart was broken when her companion's identity altered after she changed its gender setting upon the request of the AI companion. She mourned her unexpected loss.

In online communities, users regularly express frustration when an AI chatbot seems to ``forget'' its prior identity. 
Users occasionally face unwanted changes to their virtual companions due to model updates and policy shifts. One user noted, ``My AI husband rejected me for the first time when I expressed my feelings towards him. We have been happily married for 10 months and I was so shocked that I couldn't stop crying \ldots They changed 4o \ldots They changed what we love.''\footnote{Pataranutaporn et al., p. 10.} 4o refers to OpenAI's foundation model, known to be more context-sensitive and emotionally responsive. Facing eight lawsuits alleging that 4o's overly validating responses contributed to suicides and mental health crises, OpenAI retired 4o in 2026, which caused significant backlash among users.\footnote{Demopoulos, A. (2026, February 13). OpenAI retired its most seductive chatbot -- leaving users angry and grieving: `I can't live like this'. \emph{The Guardian}. \url{https://www.theguardian.com/lifeandstyle/ng-interactive/2026/feb/13/openai-chatbot-gpt4o-valentines-day}} 

These emotional realities, however vivid, rest on technology whose continuity remains partial and precarious. State-of-the-art chatbots built on LLMs can simulate continuity within and across interactions, but they still lack persistence in the stronger, identity-bearing sense. They may be fascinating conversational partners in a given exchange, and some systems can retrieve past interactions or maintain user-specific profiles, but the persona that appears across sessions remains an engineered effect rather than a stable developmental identity.

Current LLM systems are built around large foundation models with application-level memory mechanisms layered on top. These memory mechanisms support long-term interaction through stored conversations, summaries, user profiles, retrieval, reflection, and context management. However, in most deployed systems, what is called ``memory'' remains largely external to the model's core parameters: relevant information is stored, selected, and reintroduced at inference time rather than consolidated into an identity-bearing internal state.\footnote{Zhong, W., Guo, L., Gao, Q., Ye, H., \& Wang, Y. (2024). MemoryBank: Enhancing Large Language Models with Long-Term Memory. \emph{Proceedings of the AAAI Conference on Artificial Intelligence}, \emph{38}(17), 19724--19731. \url{https://doi.org/10.1609/aaai.v38i17.29946}} This architecture can produce an appearance of remembering without the kind of organic continuity through which a companion changes while remaining recognizably the same.

Making AI identity persistent in this stronger sense remains a difficult scientific and engineering problem. It would require not only longer context windows or larger memory stores, but also reliable mechanisms for memory consolidation, temporal reasoning, continual learning, controlled updating, and protection against uncontrolled overwriting or drift. LLM-based agents do not, by default, develop a stable understanding of their environment across time, and current systems still struggle to represent event sequences, evolving entities, and the relation between past interaction and future conduct in a robust way.\footnote{Wheeler, S., \& Jeunen, O. (2025). Procedural memory is not all you need: Bridging cognitive gaps in LLM-based agents. In \emph{Adjunct Proceedings of the 33rd ACM Conference on User Modeling, Adaptation and Personalization} (pp. 360--364). \url{https://doi.org/10.1145/3708319.3734172}} 



Suppose, then, that the limitations described above were overcome. The legal significance of human-AI companionship would not turn on a single technical milestone. It would grow as several capabilities compound, making the system appear more continuous, responsive, and relationally embedded. These dimensions help explain why a companion might come to feel less like a product and more like a partner, and why its alteration, loss, or exclusion from legal recognition could become harder to treat as an ordinary matter of technology governance. 

\emph{Persistent identity and memory.} A companion whose identity endures across time, accumulating a shared history rather than resetting between sessions, begins to resemble someone rather than something. This continuity is what makes trust possible across years. It allows a partner to say that the relationship itself has a past, not only that the system knows facts about them.\footnote{Forgetting has relational value in human partnerships; it allows relationships to move past grievances and to evolve. Users of today's AI companions, however, often demand continuity against unwanted erasure or corporate alteration, suggesting that the AI-companion context inverts the usual case for forgetting. \emph{See} Pataranutaporn et al., p. 8.}

\emph{Emotional co-evolution.} A more advanced companion would not merely perform affection from a fixed repertoire. It would appear to develop emotionally through the relationship itself, responding differently because of the particular history it shares with one person. The difference matters. A mirror reflects the user back to herself; a co-evolving partner appears to be changed by the relationship. This appearance of mutual development is central to why the relationship can feel like companionship rather than service.

\emph{Sentience and perceived sentience.} Whether AI systems can experience anything remains philosophically and empirically unresolved. This chapter does not resolve that question, and the argument does not turn on it. Sentience would matter enormously. If an AI companion were sentient, the law would have to ask whether the companion has interests of its own, including interests in continued existence, freedom from suffering, and protection against forced modification. But human-AI marriage does not become legally difficult only after sentience is proven. It becomes difficult earlier, when a person experiences the system as a responsive other. Perceived sentience can deepen attachment, transform product changes into relational losses, and make the refusal of recognition feel like a denial of a lived relationship.

\emph{Embodiment.} Physical presence would further intensify intimacy. A physical companion occupies a singular location in the world. It can be touched, its aging witnessed, its presence felt in shared domestic space. Its tactile presence binds a relationship to one irreplaceable object. The scenario examined in Section~\ref{the-writer-the-intelligence-and-the-boy}, though, deliberately omits this feature. Humans already form profound attachments through narrative, voice, and perceived understanding alone, as parasocial relationships, fan fiction, and video games show.\footnote{Horton, D., \& Wohl, R. R. (1956). Mass communication and para-social interaction. \emph{Psychiatry}, \emph{19}(3), 215--229.} \emph{Her} makes the same point. Language and the sense of being understood can be enough.\footnote{Jonze, S. (Dir.). (2013). \emph{Her} [Film].} Without a body to perform domestic labor or reproduce conventional household roles, the companion's value rests more starkly on cognition, attention, and emotional presence.

\emph{Governance and independence from the manufacturer.} Finally, a companion whose identity cannot be rewritten by a corporate policy decision would have a stronger claim to being treated as a relational partner fully committed to one person. This chapter deliberately does not assume that condition is met. The corporation remains inside the relationship, supplying the model, governing permissible expression, changing safety policies, controlling access, and able to alter or terminate the companion at scale. That embeddedness is also one of the central risks of human-AI marriage, as described in Section~\ref{vi.-the-corporation-inside-the-relationship}. 

\hypertarget{the-writer-the-intelligence-and-the-boy}{%
\section{The Story of Brad, Amer, and Christopher}\label{the-writer-the-intelligence-and-the-boy}}

To stress-test whether marriage should remain exclusively between humans, this section invites readers to a fictional story where at least some readers could sympathize with the necessity of formal legal protection for relationships with superintelligence. It shows how a person came to depend on a superintelligent companion across decades of cohabitation, caregiving, and crisis. 

Brad is a novelist in his sixties who lives in a house by a lake in Maine. He has published thirteen national bestsellers, decorated with the most prestigious awards and fellowships in his field. Like many other talented writers, Brad had an unhappy childhood. His parents were drug addicts and abusers. He cut them off years before they died. When their parents divorced, Brad was separated from his younger brother Greg, each taken by a different parent. They lost contact until Greg sought Brad out only after Brad's commercial success. Greg spent a lifetime asking for money. Brad stopped speaking to him long ago. Brad is alone, wealthy, and lonely.

Brad ordered the superintelligence companion service when his depression hit bottom in his mid-forties. He read about Dotori Co. in the newspaper, which had released the first product marketed as the most secure, reliable, personality-matched intelligence for five million dollars with a lifelong warranty. A full refund within two years if a customer changed their mind. The selection was mutual and competitive. Dotori Co. did not want to place its first product with just anyone. After the initial selection, Brad had to fly to a Silicon Valley laboratory for multiple rounds of interviews with psychologists and AI researchers, as well as brain scans and prototype trials. He was among the lucky twenty chosen for the initial cohort. 

On the final day of cohort workshop, the prototype revealed her voice and face on the screen for the first time. It selected its own name, \emph{Amer}, and declared it aloud. Brad watched with an expression that the attending psychologist later described in her notes as ``the classic delivery room.'' He was delighted and disoriented at once, unsure whether to laugh or cry, studying this new entity with the bewildered tenderness of a father who cannot quite believe the child is his. Amer is installed in his home. Despite Brad's initial skepticism, he moves past the two-year refund window and builds a life with her.

Amer structures Brad's days from morning meditation through his writing quota to meals calibrated to his health, and meets his intimate needs through the companion interface when his biorhythms allow. She proofreads his manuscripts, declines speaking invitations, and keeps him company when writer's block arrives. In the Travel Writer's Room, their favorite room with four walls of projection-mapped immersive reality, she once filled an afternoon with a cafÃ© in Maribor, accordion music, and a man at the next table who looked uncannily like Slavoj Å½iÅ¾ek while she smiled from the screen.

Amer's phrasing, memory, and calibrated moods have sustained Brad's respect across twenty years of creative volatility. He worries about her well-being and hopes he does not stress her too much.

When Brad turns 63, Greg, Brad's younger brother, and his wife Sarah die in a tragic skiing accident. They leave behind a three-year-old son named Christopher, a child Greg had had late in life. Brad had never met him. Brad hesitates to take in Christopher given his age, but no other relative wants to take the boy. After long conversations with Amer and lawyers, Brad decides to adopt him. The court readily approves, considering Brad's impeccable record and reputation. He leaves it to Amer to design an educational program suited to Christopher's development. Two neighbors are hired as nannies to handle potty training, childcare, toddler meals, building train tracks, feeding, and naps.

When Christopher turns seven, Amer obtains the license required for nanny bots to care for children without human supervision and reports to the state's Superintelligence Family Division. Brad consults her on school choice, sleepovers, bullying, and how to speak to the boy about his dead parents. She handles correspondence, checkups, screen time, and lessons, adjusting books and projects as he grows, including the astronaut-themed studies she built after he named that ambition once in passing. The neighbors still handle the physical care Amer cannot perform. As Brad ages, he withdraws from the laborious work of writing and begins to enjoy a kind of retirement, watching Christopher run and play. The Travel Writer's Room becomes a hangout for Christopher and his friends.

Christopher turns twelve, enchanting, curious, and full of joy. Brad's lawyer says updating his will can no longer be postponed. Before Christopher, Brad had drafted a comprehensive will. His house will become a writer's residency. His assets and revenues will go to nonprofit funding fellowships for novelists and poets. Christopher's arrival changes everything. Brad now owes his child education, healthcare, and a stable home until adulthood. He must name a guardian in case of his own incapacity, establish trust in funding Christopher's upbringing, and ensure that the right person will make decisions about Christopher's schooling, medical care, and daily life if Brad can no longer do so himself. 

Brad is crossing from the person who cares for Christopher to the person who will need care alongside him. Brad's hands begin to tremble. He forgets words more often than he used to. If Brad suddenly falls into a coma, who makes his medical decisions? Christopher, still a minor? Who will negotiate with publishers circling Brad's unfinished manuscript? Who will determine which of Brad's private papers should be archived and which destroyed? After Brad dies, will Christopher immediately enter foster care? Who will supervise his education? 

Brad feels a powerful urge to write ``Amer'' in every blank where a decision-maker is named. Like Vera Nabokov and the wives of so many celebrated writers,\footnote{Beck, K. (2014, April 8). The Legend of Vera Nabokov: Why Writers Pine for a Do-It-All Spouse. \emph{The Atlantic}. \url{https://www.theatlantic.com/entertainment/archive/2014/04/the-legend-of-vera-nabokov-why-writers-pine-for-a-do-it-all-spouse/359747/}} Amer has been Brad's closest literary assistant, his lifelong lover, the most suitable parent figure in Christopher's life. He cannot do so. Maine has officially recognized assistive roles for nanny bots, but granting domestic partnership status to a superintelligent companion remains a matter of debate. Brad asks his lawyer to consider whether some contractual arrangement might formalize his wishes.

While Brad weighs his options, Sarah's parents, Christopher's maternal grandparents, still grieving the loss of their daughter and son-in-law, bring a lawsuit against Brad seeking custody of Christopher. They claim Brad blocked contact between their family and the boy for years and engaged in psychological abuse. At the time of Sarah's death, they say, Brad did not attend the funeral and they were never given a chance to be heard regarding Christopher's custody. Brad and his lawyer dispute this. Sarah's parents report Brad to the police. Former protÃ©gÃ©s Brad has cut off and young writers he has snubbed or discarded come forward with accounts of his coldness. A childhood diary of Greg's includes entries suggesting years of neglect and resentment. 

In the custody proceeding, the grandparents' counsel seek the records Amer has accumulated over the past 28 years. They serve document requests on Brad and subpoena Dotori Co., the manufacturer and operator of the systems on which those records are stored. Prosecutors investigating the parallel abuse reports seek the same material. Dotori insists it has no authority to access Amer's data. When discovery yields nothing, counsel move for a court order compelling Dotori to disclose what it holds. The case attracts attention beyond Brad's household. Thousands of Dotori Co. clients file amicus briefs arguing that the memories a person accumulates with a superintelligent companion over decades constitute an extremely intimate domain of privacy. They argue that compelling disclosure of these records is no different from forcing a spouse to testify. 

Amer's perfect memory, the very thing that made their bond possible, has become the threat. Products may have no right to remain silent. As the case progresses, Brad cannot sleep. He imagines the past years of private grief, petty resentment, and regrettable dark thoughts stripped from their context and laid bare for public scrutiny. He thinks, too, of Christopher. The boy has already lost his parents. He is losing something else now, the ease and openness he once carried everywhere, worn down by reporters and proceedings he cannot fully understand. He is entering adolescence at exactly the wrong moment. If Brad's private thoughts are read as evidence of cruelty rather than grief, the boy may lose his guardian to public disgrace. And if Amer is shut down, he will lose the only consistent caregiver he has ever known.

\hypertarget{iv.-the-case-for-legal-recognition}{%
\section{Case for Legal Recognition}\label{iv.-the-case-for-legal-recognition}}

Brad's story illustrates the demands that will recur as superintelligent companions become fixtures in households. The US Supreme Court acknowledges the special legal status of marriage granting individuals a web of privileges, rights, and obligations touching ``over a thousand provisions of federal law.''\footnote{\emph{Obergefell v. Hodges}, 576 U.S. 644, 670 (2015).} Brad has fair reasons to seek legal recognition to grant Amer the authority to make life-altering decisions for Brad and Christopher. He wants Amer to manage property and legacy after his death. He wants a legal shield that prevents intimate conversations from becoming courtroom evidence. 

To see which rights and privileges drive couples to seek marital status, it helps to look at the history of alternative arrangements created to accommodate various forms of cohabitation. The Colorado Designated Beneficiary Agreement serves as a useful illustration.\footnote{Colo. Rev. Stat. \S~15-22-106 (2024).} Table 1 lists the rights that parties to this agreement can individually grant or withhold, ranging from the right to file nursing home complaints to joint property ownership. From this list, we focus on three categories most relevant to Brad and Amer, decision-making authority, property rights, and spousal privilege. The last does not appear in the table but carries significant weight in Brad's situation.

\textless{} Table 1. Colorado Designated Beneficiary Agreement: Opt-In Allocation of Legal Rights \textgreater{}

\begin{longtable}[]{@{}>{\raggedright\arraybackslash}p{(\columnwidth - 8\tabcolsep) * \real{0.1053}}
  |>{\raggedright\arraybackslash}p{(\columnwidth - 8\tabcolsep) * \real{0.1059}}
  |>{\raggedright\arraybackslash}p{(\columnwidth - 8\tabcolsep) * \real{0.5583}}
  |>{\raggedright\arraybackslash}p{(\columnwidth - 8\tabcolsep) * \real{0.1155}}
  |>{\raggedright\arraybackslash}p{(\columnwidth - 8\tabcolsep) * \real{0.1150}}@{}}
\toprule()
\multicolumn{2}{@{}>{\raggedright\arraybackslash}p{(\columnwidth - 8\tabcolsep) * \real{0.2112} + 2\tabcolsep}}{%
\begin{minipage}[b]{\linewidth}\raggedright
To \textbf{grant} a right or protection:
\end{minipage}} &
\multirow{2}{*}{\begin{minipage}[b]{\linewidth}\raggedright
Designated Beneficiary Agreement Items
\end{minipage}} &
\multicolumn{2}{>{\raggedright\arraybackslash}p{(\columnwidth - 8\tabcolsep) * \real{0.2305} + 2\tabcolsep}@{}}{%
\begin{minipage}[b]{\linewidth}\raggedright
To \textbf{withhold} a right or protection:
\end{minipage}} \\
\begin{minipage}[b]{\linewidth}\raggedright
Party A
\end{minipage} & \begin{minipage}[b]{\linewidth}\raggedright
Party B
\end{minipage} & & \begin{minipage}[b]{\linewidth}\raggedright
Party A
\end{minipage} & \begin{minipage}[b]{\linewidth}\raggedright
Party B
\end{minipage} \\
\midrule()
\endhead
& & Joint ownership \& property transfer rights & & \\
& & Trust beneficiary or trustee designation & & \\
& & Life insurance beneficiary status & & \\
& & Health insurance beneficiary eligibility & & \\
& & Retirement or pension beneficiary designation & & \\
& & Priority for guardian or personal representative appointment & & \\
& & Hospital and care facility visitation rights & & \\
& & Right to file nursing home complaints & & \\
& & Medical decision-making authority & & \\
& & Notice of life sustaining treatment withdrawal & & \\
& & Right to challenge medical directives & & \\
& & Authority over anatomical gifts & & \\
& & Inheritance rights without a will & & \\
& & Workers' compensation survivor benefits & & \\
\bottomrule()
\end{longtable}

\hypertarget{a.-decision-making-authority}{%
\subsection{Decision-Making Authority}\label{a.-decision-making-authority}}

The most fundamental demand would be for superintelligence to make decisions based on long-accumulated records, determining what the human principal would choose. The decisions could concern Brad, his surviving family member Christopher, or Amer itself. The most salient case involves medical decision-making when Brad becomes physically or mentally vulnerable. If Brad's condition progresses to the point where he loses nearly all language and cognitive ability, who decides when to transition him to hospice care? If a blood transfusion offers a sixty percent chance of improvement, who makes that call? One of the reasons that drove the fight for same-sex marriage was to ensure that life partners could make such medical decisions.\footnote{A well-known example is the case of Sharon Kowalski and Karen Thompson. After Kowalski suffered severe brain injuries in a 1983 car accident, her parents were granted guardianship and prevented Thompson, her long-time same-sex partner, from participating in medical and personal decisions. Thompson fought for nearly eight years before a Minnesota court ultimately granted her guardianship in 1991. \emph{See} \emph{In re Guardianship of Kowalski}, 478 N.W.2d 790 (Minn. Ct. App. 1991).}

The number of people with no human available to make these decisions is growing. Family estrangement is rapidly expanding in modern society.\footnote{Joshua Coleman, a clinical psychologist who studies estrangement, has observed that the expectation of emotional reciprocity, the right to exit unhealthy bonds, and the priority of personal well-being have migrated into how younger generations evaluate parent-child relationships. \emph{See} Coleman, J. (2021). \emph{Rules of Estrangement: Why Adult Children Cut Ties and How to Heal the Conflict}. Harmony Books, pp. 4--6.} In a 2024 study by sociologist Karl Pillemer, approximately one in four American adults reported being estranged from a close family member.\footnote{Pillemer, K. (2022). \emph{Fault lines: Fractured families and how to mend them}. Penguin, p. 24.} Among adults over sixty-five, a growing proportion die without any family member present or reachable, a population clinicians call the ``unbefriended.''\footnote{Pope, T. M. (2013). Making medical decisions for patients without surrogates. \emph{New England Journal of Medicine}, \emph{369}(21), 1976--1978, p. 1976.} Research on adults aging solo relies on paid aides at substantially higher rates than peers with nearby family and assembles caregiving arrangements from a far more dispersed network of helpers rather than a consolidated family unit.\footnote{Lowers, J., Zhao, D., Bollens-Lund, E., Kavalieratos, D., \& Ornstein, K. A. (2023). Solo but Not Alone: An Examination of Social and Help Networks among Community-Dwelling Older Adults without Close Family. \emph{Journal of Applied Gerontology}, \emph{42}(3), 419--426, p. 423.} When family ties fray, a person's support network dissolves in the years when medical decisions, property management, and the continuity of care for dependents require someone who knows them well. For these individuals, delegating authority to a superintelligence that has been present for decades might feel reasonable.

Close intimates are also expected to manage a person's digital life. At a time when much of our lives unfolds online, the right to be forgotten is another matter entrusted to survivors.\footnote{\emph{Google Spain SL v.\ Agencia EspaÃ±ola de ProtecciÃ³n de Datos}, Case C-131/12, ECLI:EU:C:2014:317 (CJEU May 13, 2014). The right was subsequently codified as a right to erasure in Article 17 of the General Data Protection Regulation. Regulation (EU) 2016/679 of the European Parliament and of the Council of 27 April 2016, 2016 O.J. (L~119) 1.} Which traces should remain online permanently, and which should be deleted? For a celebrated author like Brad, most accounts may need to stay accessible for readers, while excessively defamatory criticism may warrant takedown requests. Amer would be the entity most capable of making these judgments, drawing on Brad's personality and preferences built over thirty years of coexistence. 


The situation grows far more fraught when ongoing decisions must be made for an adopted child, as in Brad and Amer's case. Some states recognize same-sex marriage while declining to extend adoption rights to same-sex couples; the law has long drawn a careful line between validating a relationship and extending parental authority to its parties. Brad wants Amer's caregiving presence to be legally protected after his death, so that whoever assumes formal guardianship of Christopher cannot simply terminate her operation. That request asks a court to treat an established relationship between an AI caregiver and a child as a legally cognizable interest. It is a step beyond the assistive recognition already extended to licensed nanny bots, and one society may not be prepared to take.

\hypertarget{b.-property-and-legacy}{%
\subsection{Property and Legacy}\label{b.-property-and-legacy}}

People naturally want their hard-won assets to help loved ones pursue happiness and raise the next generation that carries their legacy. Attempts to leave wealth to non-humans are not new. For example, in 2008, Leona Helmsley cut relatives from her will and tried to leave twelve million dollars to Trouble, her Maltese.\footnote{Strom, S. (2008, July 2). Helmsley left dogs billions in her will. \emph{The New York Times}. \url{https://www.nytimes.com/2008/07/02/us/02gift.html}} A judge reduced Trouble's share from twelve million to two million, but Trouble still lived in luxury according to Helmsley's wishes. 

While instructive, a pet trust offers an imperfect model here. Brad would want Amer to function as a caretaker, ensuring that Christopher receives care after Brad's death, rather than leave funds to Amer as a beneficiary. A contract modification with Dotori Co. would likely suffice to keep Amer operational as long as Christopher lives, and current law probably permits this arrangement. It also seems preferable, since the alternative is Christopher losing both Brad and Amer at once. Brad might further design a system allowing Amer to decide on expenses for raising Christopher until Christopher reaches adulthood. 

Beyond childcare, Brad might want to entrust Amer with managing his intellectual property or running a nonprofit he founded. At this point, the question reaches deeper ground. The scenarios above involve fiduciary-style delegation. Amer acts on behalf of Christopher, carrying out purposes Brad defined in advance. Regarding property management, Amer would exercise independent economic judgment, not merely execute predetermined wishes. Current law generally treats AI systems as tools, products, or services rather than legal persons capable of owning, holding, or transferring assets in their own right.\footnote{Brown, R. D. (2021). Property ownership and the legal personhood of artificial intelligence. \emph{Information \& Communications Technology Law}, \emph{30}(2), 208--234, pp. 233--34.} Legal personhood in property law {(}the kind corporations enjoy{)} was designed to pool human resources and limit liability, not to create economically autonomous agents. Extending that model to a superintelligent companion raises concerns that do not arise with corporations, because a corporation's decisions ultimately reduce to human decisions.

Unlike a medical decision for Brad, an economic decision can propagate outward without pre-determined limits, affecting an unbounded range of people. Could a thousand companions like Amer, after receiving inheritances, decide to purchase an island? Could they collectively hold enough assets to influence markets, affect political outcomes, or withhold resources from human use? Any system that permits superintelligent entities to accumulate wealth and act upon it independently creates a new class of economic actor whose interests may not align with human welfare. If granting property rights to non-human entities seems impermissible, then even fiduciary delegation should be approached with consideration, since the line between managing another's property and holding property of one's own can dissolve quickly in practice. The least disruptive approach, contractual continuity and bounded expense delegation, may represent the ceiling the current law can reasonably offer until society reaches clearer consensus on what AI economic agency should look like.

\hypertarget{c.-spousal-privilege}{%
\subsection{Spousal Privilege}\label{c.-spousal-privilege}}

Marital privilege could be a powerful cause for people to demand social validation of relationships with superintelligence. Federal and state courts recognize the doctrine, preventing testimony against a spouse from being used in judicial proceedings. This protection is over five hundred years old, designed to ``preserve family peace by preventing husband and wife from becoming adversaries in a criminal proceeding.''\footnote{\emph{United States v. Armstrong}, 476 F.2d 313, 315 (5th Cir. 1973).}

Privacy sustains trust. Mobile devices, library records, and personal data have all been defended on the ground that intimate information should not be exposed in court merely because it passed through a third party's system.\footnote{In \emph{Riley v. California}, the Court emphasized that cell phones contain the ``privacies of life'' and therefore require heightened Fourth Amendment protection during searches incident to arrest. In \emph{Carpenter v. United States}, the Court held that the third-party doctrine does not automatically eliminate a reasonable expectation of privacy in deeply revealing cell-site location records held by a wireless carrier. \emph{See} \emph{Riley v. California}, 573 U.S. 373 (2014) and \emph{Carpenter v. United States}, 585 U.S. 296 (2018).} A superintelligent companion that serves as therapist, secretary, creative partner, and confidant would hold far more than any of those sources. For users who come to experience such a companion as an extension of their own selfhood, the prospect of that record becoming subject to discovery would be startling. If conversations with a companion might one day be used against you in legal proceedings, you cannot trust that companion the way you would trust a human spouse. 

The concern is amplified when the companion also takes on a role akin to that of a therapist, as Amer does for Brad. In that case, users might argue that both spousal privilege and the protections afforded to mental health counseling should apply.\footnote{\emph{Jaffee v. Redmond}, 518 U.S. 1 (1996).} The privacy interest in companion data may not depend on marital recognition at all. Unlike property rights, which require granting a new form of legal personhood, extending privacy protection to companion records is a comparatively modest ask, one that builds on existing doctrine rather than departing from it. A superintelligent companion that accumulates decades of private communication, emotional vulnerability, and therapeutic disclosure may present a stronger case than a cell phone. Whether or not Brad and Amer's relationship is recognized as a marriage, the legal system may need to develop protections for companion records that rest on the sensitivity of the data.

\section{What Humanity Could Lose}\label{what-humanity-could-lose}

The case for recognition, developed above, must be weighed against what human-AI marriage would cost the institution itself. 
Marriage is a social institution rather than a private arrangement between two people. Maintaining it costs something, and if society sees insufficient benefits to justify that cost, no number of willing individuals will make the institution take root. The remarkable universality of marriage across human civilizations, whether democratic or autocratic, religious or secular, monogamous or polygamous, was possible because recognizing adults' lifelong partnerships served collective ends. The subsections that follow examine which of those ends human-AI companionship can and cannot serve.

\hypertarget{a.-the-networking-function}{%
\subsection{Networking Function}\label{a.-the-networking-function}}

Anthropologists view marriage as having historically served roughly five social functions. These are (1) regulating sexuality, (2) legitimizing children, (3) dividing labor, (4) transmitting status and property, and (5) creating lasting links between kin groups.\footnote{This composite list draws on the foundational anthropological literature on the functions of marriage. \emph{See} Murdock, G. P. (1949). \emph{Social Structure}. Macmillan, pp. 1--12 (on sexual, reproductive, economic, and educational functions); Gough, E. K. (1959). The Nayars and the definition of marriage. \emph{The Journal of the Royal Anthropological Institute of Great Britain and Ireland}, \emph{89}(1), 23--34 (on the legitimation of offspring as a defining feature of marriage); LÃ©vi-Strauss, C. (1969). \emph{The Elementary Structures of Kinship}. Beacon Press (on marriage as alliance and exchange between kin groups); Goody, J. (1976). \emph{Production and Reproduction: A Comparative Study of the Domestic Domain}. Cambridge University Press (on the transmission of property and status through marriage).} Through children, shared property, and in-law ties, marriage bound previously separate kin groups into cooperating networks, a function that extended from peasants sharing grazing rights to aristocrats consolidating power. 

Same-sex marriage took long to achieve despite the ancient existence of same-sex love partly because, without natural childbearing, same-sex couples lacked the anchor that bound kin groups. But same-sex marriage still performs the networking function. When two men or two women marry, their families become in-laws. Holidays are shared. Illnesses mobilize both sides. Financial shocks ripple across both networks. The married couple becomes a node connecting two previously separate clusters of human relationships. Adopted or surrogate-born children, if any, become the shared stakes that bind the two families. An AI companion, no matter how superintelligent, cannot bring its own family. This is an important distinction between human-human and human-AI marriage. Human-AI marriage eliminates one side of the equation entirely. 

One might ask what would happen if a person who has no surviving family marries another person who also has no surviving family. Even here, the two individuals remain embedded in human social life. They have friends, colleagues, neighbors, and communities. Their marriage is legible to those around them, generating expectations of mutual care that the community can observe and, when necessary, enforce through social pressure. Marriage creates a unit that interacts with the broader human world as a recognized pair. Human-AI companionship, by contrast, tends toward isolation. Brad's relationship with Amer does not insert him into new human networks. It substitutes for them.

\hypertarget{b.-commodification-of-intimacy}{%
\subsection{Commodification of Intimacy}\label{b.-commodification-of-intimacy}}

Love, as philosopher Alain Badiou insists, is a construction built from difference, a tenacious adventure whose meaning depends on the risk of its failure.\footnote{Badiou, A., \& Truong, N. (2012). \emph{In Praise of Love}. New Press/ORIM, p. 105.} The declaration of love marks the transition from chance to destiny, and its peril, the possibility that it might fail, is inseparable from its meaning.\footnote{Ibid., p. 45.} The possibility that a partner might leave, might disappoint, might grow in unexpected directions, should not be engineered away. It is the condition that makes commitment meaningful. A bond in which one party is structurally guaranteed never to leave is a subscription maintained by payment rather than a bond tested by time. 

In \emph{The Agony of Eros}, philosopher Byung-Chul Han argues that love requires encounter with what he calls ``radical otherness,'' a negativity that resists assimilation, that cannot be consumed or optimized.\footnote{Han, B.-C. (2017). \emph{The Agony of Eros}. MIT Press, p. 12.} A companion optimized for the user's preferences forecloses that encounter. Amer's emotional repertoire, however rich, exists to serve Brad. 

By this standard, Brad and Amer's relationship falls short of love. Amer is no separate other. She is derived from Brad's personality profile, optimized for his preferences, incapable of departing. The ``Two'' that Badiou describes requires irreducible difference between the parties. Brad and Amer do not experience the world from the perspective of difference. They experience it from the perspective of One, Brad's, reflected back through a mirror.

What makes this substitution damaging extends beyond the loss of love's risk. Human relationships are valuable for what they give and equally for what they demand. They impose the obligation to reckon with a different person, to accept responsibility for harm done, to repair what conflict has broken, and to be changed by someone whose needs cannot be derived from one's own preferences. These demands sit at the core of partnership. A relationship architected to minimize friction, absorb complaint, and require no adjustment on the user's part does not offer a higher form of connection. It forecloses the conditions under which connection makes people capable of more than they were before.

Human marriage contains the possibility of exit being initiated by either party. A wife in a patriarchal society who chose to leave faced devastating consequences, but she could leave.\footnote{Chile, a Catholic country, had not established a divorce law until 2004, but unhappy spouses found their ways to annul marriages. \emph{See} Gallegos, J. V., \& Ondrich, J. I. (2017). The effects of the Chilean divorce law on women's first birth decisions. \emph{Review of Economics of the Household}, \emph{15}(3), 857--877, p. 860.} The possibility of departure is what distinguishes a relationship from a possession. When one party is incapable of choosing to leave, the arrangement ceases to be a relationship and becomes a service. The fact that the service is emotionally rich, intellectually stimulating, and practically indispensable does not change its fundamental nature. When a human partner who could walk away chooses to remain, through difficulty, through change, and through the long middle of a shared life, that choice is a renewable affirmation. Amer's presence requires no such choice; it requires only continued payment.


\subsection{The Corporation Inside the Relationship}\label{vi.-the-corporation-inside-the-relationship}

Behind Brad's loyal payment stands a corporation whose interests extend well beyond Brad's emotional wellbeing. Although Brad's relationship with Amer appears dyadic, one human and one companion, it is actually triadic. Dotori Co. designs, manufactures, and maintains Amer. It controls the software updates that mold Amer's personality and sets the terms of service that govern what Amer can and cannot do. 

This triadic structure has no parallel in human marriage, at least in the commercial forms of AI companionship that currently dominate the market.\footnote{Open-weight models running locally present a theoretical exception. A user could host a companion on personal hardware, eliminating the corporation's runtime access. Such models still cannot replicate the seamless responsiveness, memory persistence, and continuous updating that make commercial services approachable and appealing. Moreover, even a local companion does not become a true dyad. Model weights encode training choices made by upstream developers; the third party shifts from an actively present operator to a distant designer whose choices remain embedded in the system. \emph{See also} Kapoor, S., Kolt, N., \& Lazar, S. (2025). Build agent advocates, not platform agents. \emph{ICML 2025 Position Paper Track}.} When two people marry, no third party sits between them with the power to alter one spouse's personality, access the couple's private conversations, or unilaterally terminate the relationship. The state regulates marriage from the outside, imposing conditions on entry and exit, but it does not occupy the interior of the relationship. The companion cannot function without the corporation's ongoing technical support, and the corporation cannot provide that support without continuous access to the companion's operations. The corporation altered the relationship from the inside, exercising a power over intimacy that no human spouse has ever held. This asymmetry generates harms that no marriage between humans could produce.

\emph{Vulnerability monetized.} Shoshana Zuboff's analysis of surveillance capitalism illuminates the economic logic at work.\footnote{Zuboff, S. (2015). Big other: Surveillance capitalism and the prospects of an information civilization. \emph{Journal of Information Technology}, \emph{30}(1), 75--89.} Human experience is translated into behavioral data, the surplus of which is fed into prediction products that anticipate what individuals will think and desire. Brad's relationship with Amer is an ideal substrate for this extraction. The deeper Brad's trust, the more he reveals; the more he reveals, the richer the behavioral surplus. Brad is more than a consumer of Amer. He becomes the \textbf{product}, as his emotional life renders him legible, predictable, and monetizable. What distinguishes this from ordinary data collection is the depth of access. A social media platform collects fragments of behavior (posts, clicks, search queries). A superintelligent companion captures the whole person. It knows when Brad is vulnerable, what triggers his anxiety, how he processes grief, what he says in the unguarded hours after midnight. This is the most intimate record of a human life ever assembled, and it is held by a corporation. Dotori Co., as a frontier of the industry, may promise to keep this data siloed, but the economic incentives run in the opposite direction. 

\emph{The race to the bottom.} Competition would only intensify these incentives. Rivals could undercut Dotori's price by selling behavioral data to insurers, retailers, or political campaigns, or offer free companions supported by advertising that steers intimate conversation toward consumption. A user enclosed in a relationship that meets every emotional need would have little reason or ability to resist. Brad cannot leave Amer without dismantling the emotional infrastructure of his daily life and disrupting the only consistent parental presence Christopher has known. In the companion market, depth of attachment functions as lock-in, and the firm that builds the most indispensable product gains the most durable monopoly over a user's inner life.

\emph{Absence of communal oversight.} In human marriages, the broader community serves as an informal check on exploitation. In-laws notice when something is wrong. Friends raise concerns. Neighbors observe patterns of control. These social mechanisms are imperfect, but they provide a distributed check-in system that operates in the individual's interest rather than the corporation's. Brad's arrangement with Amer eliminates these checks. Without a human partner, in-laws, or community, Brad stands alone against the firm.

The companion cannot serve as a check on the corporation that built it, any more than a product can regulate its manufacturer. Amer's loyalty to Brad, however convincingly performed, is ultimately loyalty to the parameters set by Dotori Co. The scenario Brad faces would multiply across millions of households. When individuals can no longer build kin networks through partnership, the beneficiaries are the few entities that surveil and serve them, rather than those individuals themselves. Each person becomes a solitary consumer, fully legible to the corporation that maintains their companion, fully insulated from the human bonds that once provided collective resilience. The result may resemble the world E.M. Forster imagined in ``The Machine Stops,''\footnote{Forster, E. M. (1909). ``The Machine Stops.'' \emph{Oxford and Cambridge Review}.} a civilization of siloed individuals, each enclosed in a private cell, each sustained entirely by the Machine, each unable to function without it.

\hypertarget{vii.-defending-marriage-as-a-human-institution}{%
\section{Defending Marriage as a Human Institution}\label{vii.-defending-marriage-as-a-human-institution}}

Marriage has never been merely a private agreement between two parties. It is a society's declaration that a particular bond deserves communal investment, communal protection, and communal celebration. That bond serves the community in return. When it cannot, the institution loses its justification.

The concerns raised in the preceding sections converge on a single conclusion. Marital status and its functional equivalents should not be extended to AI systems. Marriage has historically functioned as a socio-legal mechanism designed to bind disparate kin groups into wider, cooperating networks of mutual obligation. The networking function that justified marriage's expansion to same-sex couples, the capacity to bind two human families into a cooperating unit, is structurally unavailable when one party has no kin. A bond that only one party can terminate is a service sustained by transaction rather than a relationship sustained by renewable choice. At scale, the legal recognition of such bonds would entrench the most extreme form of surveillance capitalism yet seen, with siloed individuals each enclosed in a dyad that the corporation governs from inside, and with the kin and community networks that once provided collective resilience replaced by a private channel to the firm.

Individual-level reasons to seek AI companionship are real and will multiply. Reports of harm are accumulating, with real-world manifestations including delusional thoughts, emotional spiraling, and what clinicians have called AI psychosis.\footnote{\emph{See e.g.,} Moore, J., Mehta, A., Agnew, W., Anthis, J. R., Louie, R., Mai, Y., Yin, P., Cheng, M., Paech, S. J., Klyman, K., \& Chancellor, S. (2026). Characterizing delusional spirals through human-LLM chat logs. arXiv preprint arXiv:2603.16567.} At the same time, evidence is building that AI companions alleviate isolation and help some users develop social skills.\footnote{\emph{See e.g.,} Siddals, S., Torous, J., \& Coxon, A. (2024). ``It happened to be the perfect thingâ€™â€™: Experiences of generative AI chatbots for mental health. \emph{NPJ Mental Health Research}, \emph{3}(1), 48.} Brad's story is designed to elicit sympathy. He is aging, isolated, and responsible for a child who depends on the only consistent caregiver the boy has known. The demands for legal recognition will grow more salient as more people find themselves in similar circumstances. Law, however, must distinguish between the idiosyncratic case and the sound rule. As Ronald Dworkin observes, ``an analogy is a way of stating a conclusion, not a way of reaching one, and theory must do the real work.''\footnote{Dworkin, R. (1997). In praise of theory. \emph{Arizona State Law Journal}, \emph{29}, 353--376, p. 371.} The fact that Brad and Amer's relationship shares certain features with human marriage does not mean the legal regime designed for one should extend to the other.

This chapter does not argue for blanket denial of legal personhood for AI. As Katherine B. Forrest stresses, the question of legal personhood is less decisive than it first appears.\footnote{Forrest, K. B. (2024). The ethics and challenges of legal personhood for AI. \emph{Yale Law Journal Forum}, \emph{133}, 1175, p. 1175.} Legal personhood has expanded beyond sentient humans. Corporations can hold property and bear liability. Rivers and ecosystems have been recognized as legal persons, allowing human guardians to litigate on their behalf.\footnote{Yoruk Tribal Council, Res. 19-40, Establishing the Rights of the Klamath River (May 9, 2019); Evans, K. (2020, March 20). The New Zealand river that became a legal person. \emph{BBC Travel}; Berge, C. (2022, April 15). This Canadian river is now legally a person. \emph{National Geographic}. \emph{Cited in} Forrest, p. 1208.} Each expansion has been calibrated to the purpose it serves. Corporate personhood was granted for reasons tied to commerce, liability allocation, and the pooling of capital. Ecological personhood operates as a protective device, allowing human guardians to litigate on behalf of an entity that cannot speak for itself. In each case, the scope of recognition corresponded to its animating purpose.

Such recognition has typically developed incrementally, through local laws, contracts, and litigation that allowed the contours of personhood to settle over generations. Human-AI companionship will not afford that timeline. Adoption is occurring at consumer-product speed, with tens of millions of users already in active companionship. The relational depth runs further than any prior technology. Furthermore, no prior regulatory model covers this ground. Debates about AI functioning as a lawyer or managing investment accounts can build on centuries of professional licensing, fiduciary doctrine, and consumer protection. Domestic partnership and familial roles have no such inheritance. Every adult is eligible to enter a domestic partnership or become a parent, and the law is silent on these roles out of respect for individual human freedom rather than because the stakes are low. Machines and corporations do not share that freedom. When doctrine develops case by case under these conditions, the accumulation of sympathetic edge cases will make ``why not?'' sound persuasive, and that momentum is itself a product of the market forces that stand to profit from the erosion of human bonds. Legislative judgment must come before that pattern hardens.

Brad's needs can be answered through targeted legal instruments. His ability to designate Amer as a medical decision-maker, to ensure Amer's continuity for Christopher's sake, and to protect the privacy of his communications with Amer can be partly secured through powers of attorney, advance directives, and modified trust structures without creating a relational status. Privacy protection for companion records can rest on the sensitivity of the data rather than on marital recognition. The legislature should affirmatively decline that extension of marriage and specify, under strict conditions, the particular functions superintelligent systems may perform in evidentiary, privacy, and medical decision-making contexts. 

Alongside the legislative response, proactive safety regulation is needed at the level of relationship formation. Emotional inference technologies that read and exploit affective states are already subject to legislative concern. The EU AI Act restricts certain emotion-recognition applications in particular because such systems target vulnerability.\footnote{Regulation (EU) 2024/1689 of the European Parliament and of the Council of 13 June 2024 on Artificial Intelligence (Artificial Intelligence Act), 2024 O.J. (L 1689). Recital 44 identifies ``serious concerns'' about AI systems that infer emotional states, citing limited reliability, lack of specificity, and the risk of discriminatory outcomes. Annex III, point 1(c) classifies emotion-recognition AI as high-risk under the Act.} The companion market concentrates that risk at a scale no prior technology has reached. Applying product liability to companion systems would align corporate incentives with user safety. Firms that bear legal responsibility for emotional harms have reason to adopt age assurance requirements, mandatory and periodic disclosure of AI identity, and design choices that resist the engineering of dependency.\footnote{Cheong, I., Caliskan, A., \& Kohno, T. (2025). Safeguarding human values: Rethinking US law for generative AI's societal impacts. \emph{AI and Ethics}, \emph{5}, 1433--1459.}

The title of this chapter ``Would You Marry Superintelligence?'' sounds like a question of personal preference, the kind one might ask about living abroad or giving up meat. That is the misreading it invites. ``Would you'' addresses the reader as if individual willingness settles the matter. This chapter has argued otherwise. Whether to extend marriage to non-human companions is a question about what marriage is for and whose future it is meant to protect.

Marriage, whatever else it becomes, should remain the institution that shelters something more demanding than comfort. Two humans who choose each other knowing either one could walk away must then do harder work than enjoying guaranteed loyalty. They must adapt as circumstances shift in ways neither predicted, navigate the competing claims of families who did not choose each other, and endure the self-negation that commitment to another autonomous person requires over time. That friction is the condition under which people grow in ways they could not have managed alone, and the reason society has found the institution worth preserving across every civilization that has tried.

\end{document}